\documentclass{paper}
\usepackage{epsfig, graphicx}
\usepackage[english]{babel}

\begin{document}
\title{Diffuse X-Ray Scattering near a Two-Dimensional Solid -- Liquid
Phase Transition at the n-Hexane -- Water Interface}
\author{\small Aleksey M. Tikhonov\/\thanks{tikhonov@kapitza.ras.ru}}
\maketitle
\leftline{\it Kapitza Institute for Physical Problems, Russian Academy of Sciences,}
\leftline{\it ul. Kosygina 2, Moscow, 119334, Russia}

\rightline{\today}

\abstract{According to experimental data on X-ray scattering and reflectometry with the use of synchrotron radiation, a two-dimensional crystallization phase transition in a monolayer of melissic acid at the n-hexane - water interface with a decrease in the temperature occurs after a wetting transition.}

\vspace{0.25in}

\large

The observation of a two-dimensional solid -- liquid
phase transition at the oil -- water interface was
reported in \cite{LeiBain, Tamam, Tikh1}.
In this work, the temperature
dependence of the intensity of diffuse (nonspecular)
scattering of 15-keV photons at the n-hexane -- water
interface, where such a transition occurs in the
adsorbed layer of melissic acid (C$_{30}$-acid) \cite{Tikh1}, is studied.
It is shown below that the diffuse scattering intensity
in the low-temperature crystal phase at the interface
is one or two orders of magnitude higher than that
in the high-temperature phase, which indicates the
existence of an extended transverse structure with a
thickness of $\sim 200$\,\AA{} in the former phase. The analysis
of experimental data within the theory of capillary
waves indicates that the two-dimensional crystallization
transition at the interface with a decrease in the
temperature occurs after the wetting transition.

Samples of macroscopically flat n-hexane -- water
interface were prepared and studied by the method
presented in \cite{s-cell, acid-c30-1} in a stainless steel cell with dimensions
of the interface of $75$\,mm$\times$\,150\,mm whose temperature
was controlled by means of a two-stage thermostat.
Systems with the volume concentration of C$_{30}$-acid in
n-hexane $c \approx 0.2$\,mmol/kg ($\approx2\times 10^{-5}$) and the
amount of material sufficient for covering of the interface
with $\sim 10^2$ monolayers of acid were studied. Saturated
hydrocarbon C$_6$H$_{14}$ with the boiling temperature
$T_b\approx 342$\,K and the density at 298\,K $\approx 0.65$ g/cm$^3$ was
preliminarily purified by multiple filtration in a chromatographic
column. A solution of sulfuric acid (pH = 2)
in deionized water (Barnstead, NanoPureUV)
was used as the lower bulk phase, where  C$_{30}$H$_{60}$O$_2$ is
hardly dissolved. The diffuse scattering intensity was
measured for a sample that was aged for no less than
12 h after a change in the temperature of the cell. In
order to prevent the formation of gas bubbles at the
interface, the sample was "annealed": the liquids in
the cell were heated to $T \cong T_b$ and were then cooled
below $T_c$.

\begin{figure}
\hspace{0.5in}
\epsfig{file=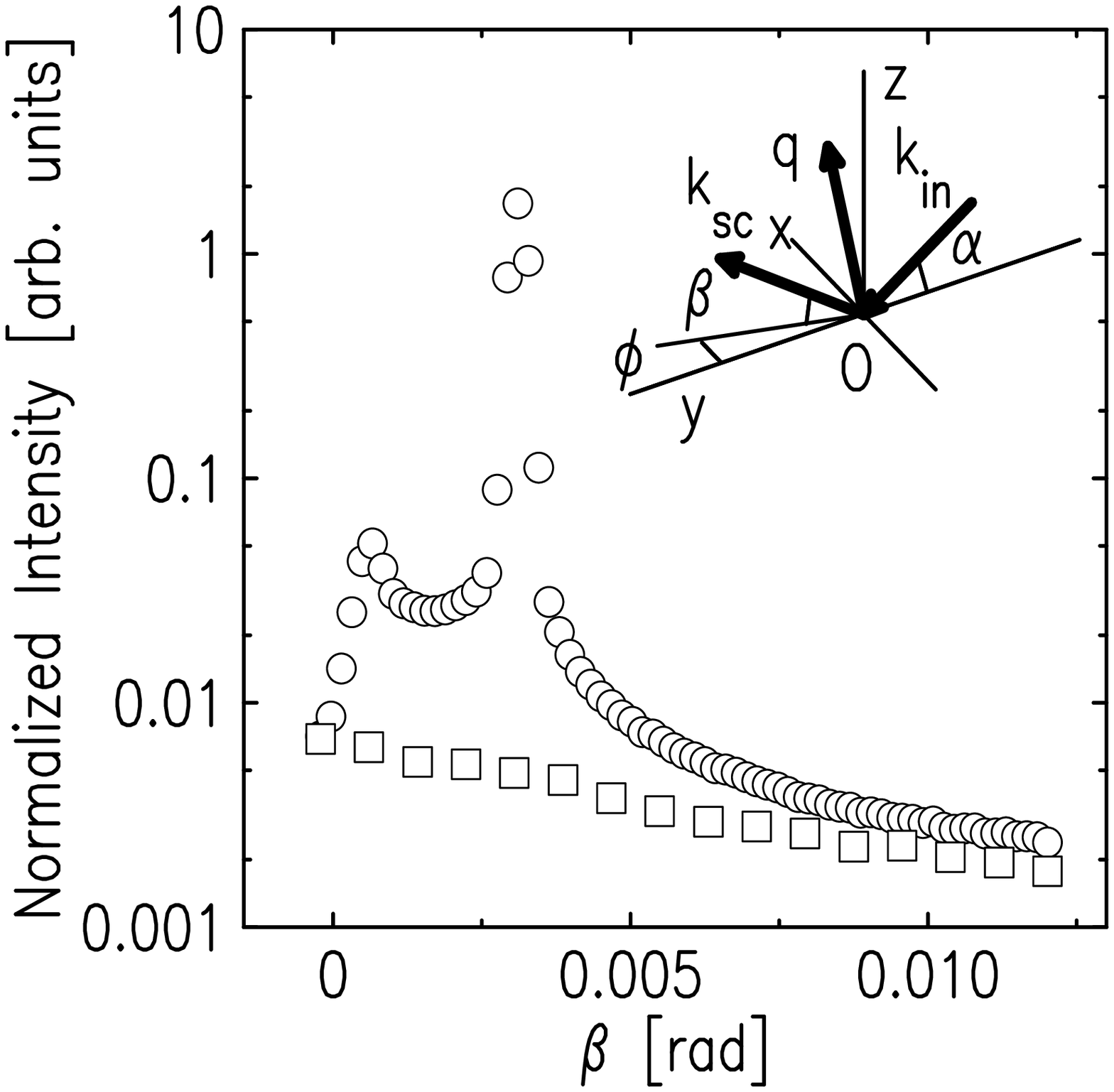, width=0.7\textwidth}

Figure 1. Angular dependencies of the scattering intensity at
the grazing angle $\alpha \approx 3.3 \cdot 10^{-3}$\,rad ($\approx 0.19^\circ$) and $T = 293.2$\,K: (circles) the total scattering intensity $I(\beta)/I_0$
and (squares) background from the scattering of the incident
beam in the bulk of n-hexane $I_b/I_0$. The inset shows
the kinematics of scattering in the coordinate system where
the $xy$ plane coincides with the boundary between the
monolayer and water, the $Ox$ axis is perpendicular to the
direction of the beam, and the $Oz$ axis is directed along the
normal to the surface against the gravitational force.
\end{figure}

The reflectometry data obtained with synchrotron
radiation previously reported for this system show that
molecules of C$_{30}$-acid are adsorbed at the n-hexane -- water interface
in form of the Gibbs monolayer with
the thermodynamic parameters ($p,\,T,\,c$) \cite{Tikh1}. A sharp
phase transition from a crystal state with the area per
molecule $A=17\pm 1$\,\AA$^2$ to a liquid state with the area
per molecule $A=23 \pm 1$\,\AA$^2$ occurs in the monolayer at
a pressure of $p=1$\,atm and a temperature of
$T_c\approx 293.5$\,K. The density of the low-temperature
solid phase of the Gibbs monolayer corresponds to the
packing in the crystal phase of Langmuir monolayer of
C$_{30}$-acid on the surface of water and is close to the volume
density of the corresponding crystal \cite{Jacquemain, Small}. The
density of the high-temperature phase is close to the
density of a high-molecular-weight hydrocarbon liquid and
corresponds to, e.g., the density of the liquid phase of
the Gibbs monolayer of melissyl alcohol at the n-hexane–
water interface \cite{Small, TAMSCH1}.

With an increase in the temperature in a close
vicinity of $T_c$ ($\Delta T <0.2$\,Ê), a significant fraction of
C$_{30}$-acid molecules adsorbed in a solid monolayer
leave the interface and are dissolved in the bulk of
n-hexane: the density of the monolayer decreases by $\approx 30\%$
and the thickness of the monolayer simultaneously
decreases by $\approx 15\%$. For both phases, a qualitative
two-layer model satisfactorily describes reflectometry
data and is in agreement with the structure of
a linear chain molecule of melissic acid C$_{30}$H$_{60}$O$_2$
with a length of $\approx 41$\,\AA{}. The formation of the first layer
involves polar head parts -COOH, whereas the second
layer is formed by hydrophobic hydrocarbon tails
-C$_{29}$H$_{59}$.

The scattering intensity $I$ at the n-hexane -- water
interface was measured by a universal spectrometer
for studying the liquid surfaces at the X19C station
of the NSLS synchrotron \cite{x19c}. In experiments, a
focused monochromatic beam with the wavelength
$\lambda=0.825 \pm 0.002$ \,\AA{} and an intensity of $ \sim 10^{10}$\,photons/s
was used. Owing to a large depth of penetration
of radiation into the hydrocarbon solvent ($\approx 19$\,mm)
and a quite high brightness of the source of synchrotron
radiation (bending magnet), scattering data can
provide information on the microstructure of the surface
layer supplementing previous reflectometry data.

In grazing geometry, the kinematics of scattering
on the macroscopically flat interface oriented by the
gravitational force is conveniently described in the
coordinate system whose origin $O$ is at the center of
the illuminated region, the $xy$ plane coincides with the
interface between the monolayer and water, the
$Ox$ axis is perpendicular to the beam direction, and the
$Oz$ axis is normal to the surface and is opposite to the
gravitational force (see Fig. 1). Let {\bf k}$_{\rm in}$
and {\bf k}$_{\rm sc}$ be the
wave vectors of the incident and scattered beams with
the amplitude $k_0= 2\pi/\lambda $ in the direction of the
observation point, respectively.
The grazing angle $\alpha << 1$
and scattering angle $\beta << 1$ lie in the $yz$ plane,
and $\phi \approx 0$ is the angle between the incident beam and
scattering direction in the $xy$ plane.
In the case of specular reflection ($\alpha = \beta$, $\phi=0$), the scattering vector
{\bf q = k$_{\rm in}$ {\rm -} k$_{\rm sc}$} is directed along the $Oz$ axis and has
the length $q_z=k_0(\sin\alpha+\sin\beta)$$\approx 2k_0\alpha$. At $\alpha \neq \beta$,
the scattering vector ${\bf q}$ has the components
$q_x=k_0\cos\beta\sin\phi$$\approx k_0\phi$ and $q_y=k_0(\cos\beta\cos\phi-\cos\alpha)$
$\approx k_0(\alpha^2-\beta^2)/2$ in the plane of interface.

When measuring the scattering intensity $I(\beta)$, the
vertical size of the incident beam with the angular
divergence $\Delta\alpha=d/l$$\approx 10^{-4}$\,rad near the surface of the
sample was $\approx 0.05$\,mm and was controlled by a pair of
collimating slits with the vertical gap $d=0.05$\,mm
spaced by the distance $l\approx 60$\,cm.
The distance from the input slit in front of
the sample to the detector was $L_1\approx 90$\,cm.
The gap in all slits in the horizontal plane
was $D\approx 10$\,mm, which was much larger than the
horizontal size of the incident beam $\sim 2$\,mm.
The intensity $I(\beta)$ was measured by a point detector with
the angular resolution in the horizontal plane $\Delta \phi = D/L_1$$\approx 10^{-2}$\,rad
and with the angular resolution in the plane of incidence
$\Delta\beta=2H_d/L_2$ $\approx 3\cdot10^{-4}$\,rad,
where $2H_d = 0.2$\,mm is the gap in the slit in
front of the detector and $L_2\approx 70$\,cm is the distance
from the center of the sample.

Below, $I_0$ is a quantity proportional to the intensity
of the incident beam, which was controlled in the
experiment immediately before the entrance of the
beam to the cell. Circles in Fig. 1 are the data on the
normalized scattering intensity $I(\beta)/I_0$ measured at
the grazing angle $\alpha \approx 3.3 \cdot 10^{-3}$\,rad ($\approx 0.19^\circ$)
and $T = 293.2$\,K. Each point is obtained by summation of
photons specularly reflected and diffusely scattered by
the surface in the illuminated region with an area of $A_0\approx 30$\,mm$^2$
at the center of the interface of the sample
in the direction $\beta$ and $I_b$ photons scattered in the
bulk of n-hexane on the path to the interface. For the
independent determination of the contribution $I_b$ in $I(\beta)$,
the experimental sample cell was
displaced down along the $Oz$ axis by $\sim 0.2$\,mm so that
the beam propagated slightly above the interface. In
this case, the detected background increased to  $\approx 2I_b$,
because the length of the path of the photon beam
in the hydrocarbon solvent increased by a factor of
about 2. The background $I_b/I_0$ thus measured is
shown by squares in Fig. 1.

\begin{figure}
\hspace{0.5in}
\epsfig{file=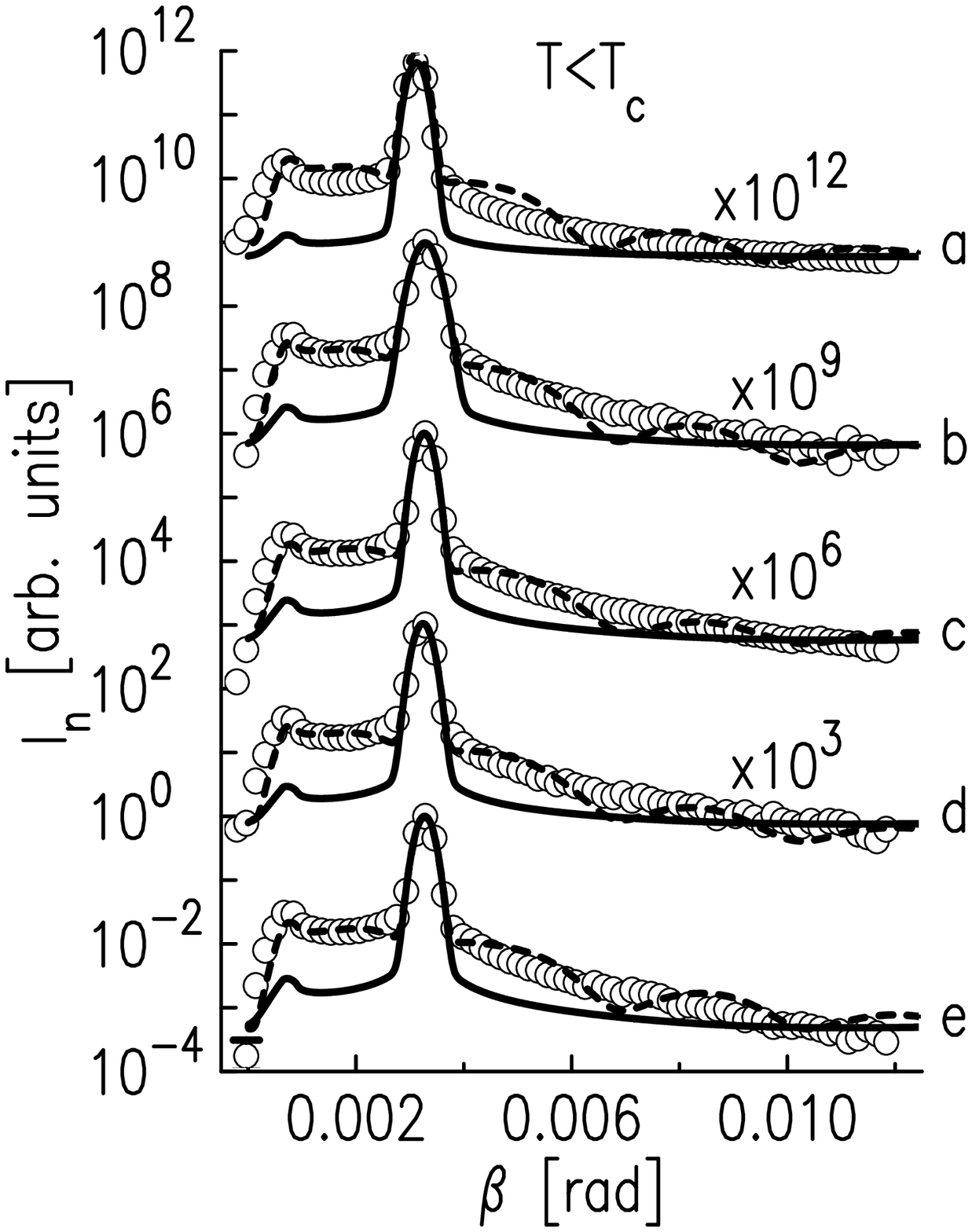, width=0.7\textwidth}

Figure 2. Angular dependences of the surface scattering
intensity $I_n$ at the grazing angle $\alpha \approx 3.3 \cdot 10^{-3}$\,rad for the
n-hexane -- water interface at various temperatures $T<T_c$:
$T =$ (a) 294, (b) 290, (c) 289, (d) 287, and (e) 285\,K. The
solid lines correspond to the monolayer model given by
Eq. (13), whereas the dashed lines correspond to the
extended-layer model specified by Eq. (14).
\end{figure}

Figures 2 and 3 show the data on the normalized
surface scattering intensity $I_n (\beta) \equiv (I(\beta)-I_b)/I_0$
(normalization condition $I_n(\alpha)\equiv 1$) obtained in the
temperature range of 285 to 335\,K. The most intense
peak on curves corresponds to the specular reflection $\beta = \alpha$,
 whereas the peak against the diffuse background
at $\beta \to 0$ corresponds to the angle of total
external reflection $\alpha_c \approx 10^{-3}$\,rad
($\approx 0.05^\circ$) \cite{Yoneda}. Scattering
occurs in the range of the characteristic in-plane
lengths $2\pi/q_y$ $\sim 10^{-5} - 10^{-6}$\,m. The long-wavelength
limit is specified by the vertical resolution of the
detector $\Delta\beta$, where the short-wavelength limit is specified
by the maximum value $\beta \sim 1.2\cdot 10^{-2}$\,rad ($\approx 0.7^\circ$)
at which the surface and bulk components in the scattering
intensity can still be separated from each other.

In the distorted wave Born approximation
(DWBA), the surface scattering intensity $I_n$ of the
monochromatic photon beam $I_0$ is the sum of diffuse
scattering $I_{\rm diff}$ and specular reflection $I_{\rm spec}$ \cite{Sinha, Holy}

\begin{equation}
I_n = I_{\rm diff} + I_{\rm spec}.
\end{equation}

Below, we examine only nonspecular scattering of
photons by thermal fluctuations of the surface of the
liquid (capillary waves), which are described by the
correlation function \cite{CW, Braslau, Schwartz, McClain, MWS}:
\begin{equation}
\langle z(0) z(r) \rangle = \frac{k_BT}{2\pi\gamma}K_0\left[\left(\frac{g\Delta\rho_m}{\gamma}(r^2+r_0^2)\right)^{1/2} \right],
\end{equation}
where $r^2=x^2+y^2$ is the square of the distance
between two points on the surface, $g$ is the gravitational
acceleration, $\gamma$ is the surface tension coefficient, $k_B$
is the Boltzmann constant, $\Delta\rho_m\approx 0.34$\,g/cm$^3$
is the difference between the densities of
water and n-hexane, $K_0(t)$ is the modified Bessel
function of the second kind, and $r_0$ is determined by
the square of the rms width of the interface
$\sigma^2_{cw} = (k_BT)/(2\pi\gamma) K_0(r_0\sqrt{g\Delta\rho_m/\gamma})$.

The averaging of $I_{\rm diff}$ over grazing angles $\alpha$ gives
\begin{equation}
I_{\rm diff}= \frac{1}{\Delta\alpha}
\int\limits_{\alpha-\Delta\alpha/2}^{\alpha+\Delta\alpha/2}\frac{1}{\sin\alpha}
\int\limits_{\Delta\Omega}
\left(\frac{ \displaystyle d\sigma}{\displaystyle d\Omega}\right)_{\rm diff}d\Omega d\alpha,
\end{equation}
where $d\Omega = \sin(\pi/2 - \beta)d\beta d\phi \approx d\beta d\phi$ and $\Delta\Omega$ is
the solid angle of photon collection by the detector.

\begin{figure}
\hspace{0.5in}
\epsfig{file=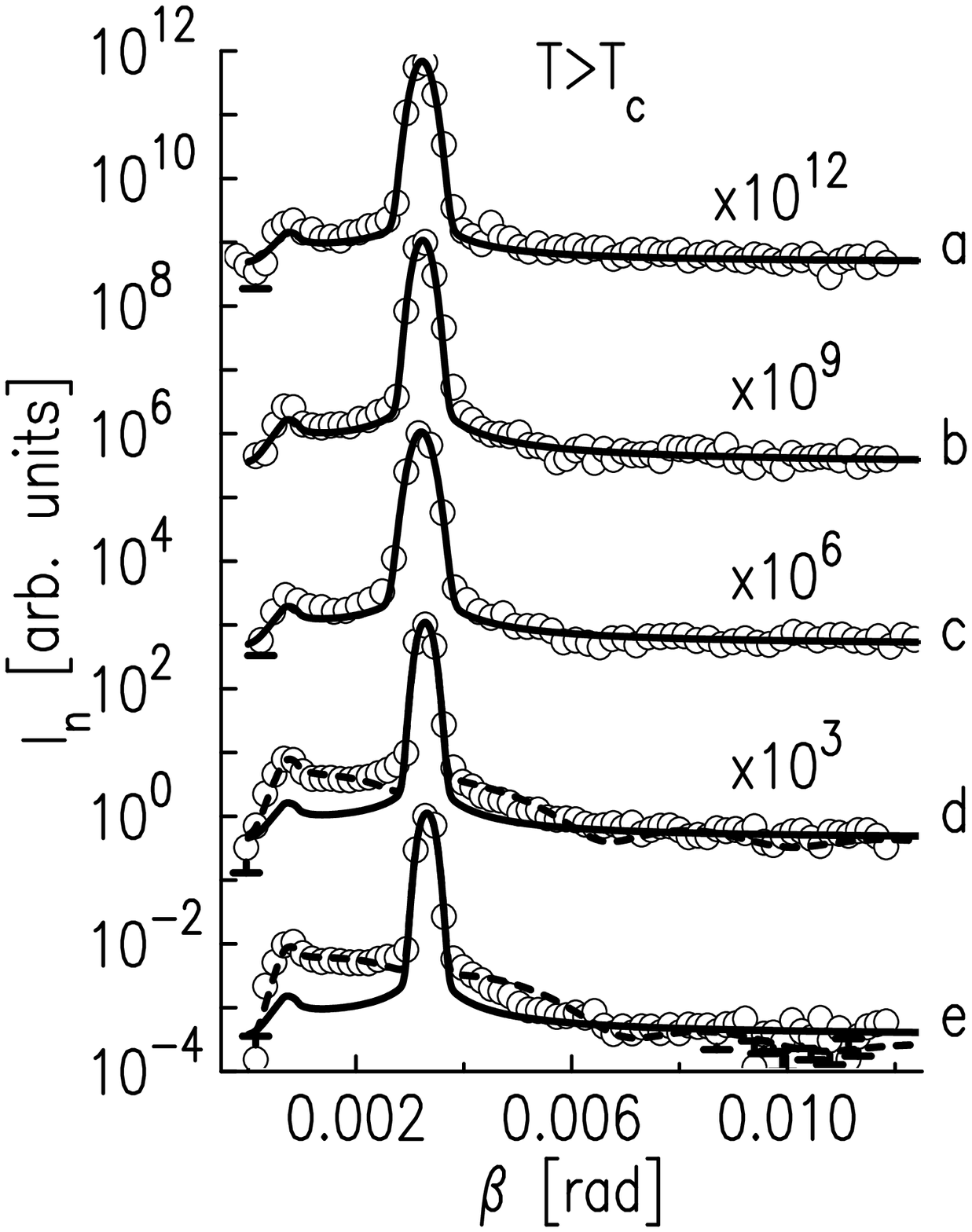, width=0.7\textwidth}

Figure 3. Angular dependences of the surface scattering
intensity $I_n$ at the grazing angle $\alpha \approx 3.3 \cdot 10^{-3}$\,rad for the
n-hexane -- water interface at various temperatures $T>T_c$:
$T =$ (a) 335, (b) 323, (c) 318, (d) 308, and (e) 298\,K. The
solid lines correspond to the monolayer model given by
Eq. (13), whereas the dashed lines correspond to the
extended-layer model specified by Eq. (14).
\end{figure}

The differential cross section for diffuse scattering is
given by the expression \cite{Sinha}
\begin{equation}
\left(\frac{\displaystyle d\sigma}{\displaystyle d\Omega}\right)_{\rm diff} =\frac{\displaystyle q^4_c}{\displaystyle (16\pi)^2}
|T(\alpha)|^2|T(\beta)|^2|\Phi(\sqrt{q_zq_z^t})|^2S({\bf q}^t)
\end{equation}
where the $z$ component of the scattering vector in the
lower phase has the form
\begin{equation}
q_z^t = \frac{2\pi}{\lambda}\left[(\alpha^2-\alpha^2_c)^{1/2}+(\beta^2-\alpha^2_c)^{1/2}\right].
\end{equation}
The angle of total external reflection
$\alpha_c$ ($q_c=2k_0\sin\alpha_c$) is related to the difference
$\Delta\rho\approx0.11$\,{\it e$^-$/}{\AA}$^3$ between the volume electron densities of
n-hexane ($\rho_h\approx0.22$\,{\it e$^-$/}{\AA}$^3$) and water ($\rho_w\approx0.33$\,{\it e$^-$/}{\AA}$^3$): $\alpha_c =\lambda\sqrt{r_e\Delta\rho/\pi}\approx 10^{-3}$\,rad, where
$r_e =2.814\cdot10^{-5}$\,\AA{} is the classical radius of the electron.
In Eq. (4), $T(\theta)$ the Fresnel transmission coefficient
for the amplitude of the wave with the polarization
of synchrotron radiation on the plane of the interface
is given by the formula
\begin{equation}
T(\theta)=\frac{2\theta}{\theta + (\theta^2 - \alpha^2_c)^{1/2}},
\end{equation}
and the structure factor of the interface,
\begin{equation}
\Phi(q)=\frac{1}{\Delta\rho}\int^{+\infty}_{-\infty}
\left\langle\frac{d\rho(z)}{dz}\right\rangle e^{iqz} dz,
\end{equation}
is determined by the Fourier transform averaged over
the illuminated area $A_0$ of the derivative of the electron
density distribution $\rho(z)$ along the $Oz$ axis. The last
factor in Eq. (4) has the form
\begin{equation}
S({\bf {q}}^t) \approx \int\int_{A_0}\langle z(0) z(r) \rangle e^{iq_{x}x + iq_{y}y}dxdy
\end{equation}
at $q^2_z \langle z(0) z(r) \rangle <<1$.

The substitution of Eq. (4) into Eq. (3) and the
subsequent integration over the variables $\phi$ and $x$
simplify the two-dimensional Fourier transform $\langle z(0) z(x,y) \rangle$
to a one-dimensional Fourier transform
in variable $y$. Further, using the relation $F_t[K_0(|t|)](\omega)=\sqrt{\pi/(2\omega^2+2)}$
for the Fourier transform,
we obtain
\begin{equation}
\begin{array}{l}
I_{\rm diff} \approx \frac{\displaystyle \lambda q^4_c }{\displaystyle 512\pi^2}
\frac{\displaystyle k_B T}{\displaystyle \Delta\alpha\gamma}\times
\\ \\
\times
\displaystyle\int\limits_{\alpha-\Delta\alpha/2}^{\alpha+\Delta\alpha/2}
\int\limits_{\beta-\Delta\beta/2}^{\beta+\Delta\beta/2}
\frac{\displaystyle |T(\alpha)|^2|T(\beta)|^2|\Phi(\sqrt{q_zq_z^t})|^2
}{\displaystyle \alpha\sqrt{q^2_y+g\Delta\rho_m/\gamma}} d\beta d\alpha
\end{array}
\end{equation}
The further integration in Eq. (9) is performed numerically.

The intensity of specular reflection in Eq. (1) is
given by the expression
\begin{equation}
I_{\rm spec}=f(\alpha, \beta)R(\alpha),
\end{equation}
where the reflection coefficient
\begin{equation}
R(\alpha)=\left|\frac{\displaystyle q_z-q_z^t}{\displaystyle q_z+q_z^t}\right|^2\left|\Phi(\sqrt{q_zq_z^t})\right|^2,
\end{equation}
is calculated at $\alpha \equiv \beta$ with the use of Eqs. (5) and (7).

The instrumental angular resolution function $f(\alpha, \beta)$,
which includes the Gaussian distribution of
the intensity of the beam in the plane of incidence, has
the form \cite{MWS}
\begin{equation}
f(\alpha, \beta)=\frac{1}{2}\left[{\rm erf}\left(\frac{H+H_d}{\sqrt{2}L_1\Delta\alpha}\right) - {\rm erf}\left(\frac{H-H_d}{\sqrt{2}L_1\Delta\alpha} \right)\right],
\end{equation}
(where $H=(\beta - \alpha)L_2$ and the error function ${\rm erf}(t)=(2/\sqrt{\pi})\int_{0}^{t}{e^{-s^2}ds}$) provides better agreement with experiments than the simplest trapezoidal resolution
function \cite{Braslau}.

On one hand, according to experimental data
shown in Fig. 2, the diffuse background at small $\beta$ values
in the solid phase reaches $\approx 5 \cdot 10^{-2}$ of the height of
the specular reflection peak and hardly depends on the
temperature up to $T_c$ at which $R$ changes stepwise. On
the other hand, the data shown in Fig. 3 indicate a
gradual decrease in the scattering intensity $I_{\rm diff}$ to $\approx 2 \cdot 10^{-3}$
in the liquid phase with an increase in the
temperature from $T_c$ to $T^*\approx 320$\,K.

In \cite{Tikh1, acid-c30-1}, the phases of the monolayer of melissic
acid were described within a qualitative two-layer
model with the structure factor (7) of the form
\begin{equation}
\Phi(q)_m = \frac{e^{-\sigma_R^2q^2/2}}{\Delta\rho}\sum_{j=0}^{2}{(\rho_{j+1}-\rho_j)e^{-iqz_j}},
\end{equation}
where $z_0=0$, $\rho_0=\rho_w$, and $\rho_3 = \rho_h$. For the solid
phase, the electron densities are $\rho_1 \approx 1.16\rho_w$ and $\rho_2 \approx 1.02\rho_w$
and the coordinates of the interfaces
between the layers are $z_1 \approx 15$\,\AA{} and $z_2 \approx 41$\,\AA{}(the
length of the C$_{30}$-acid). In the liquid phase of the
monolayer, $\rho_1 \approx 1.1\rho_w$, $\rho_2 \approx 0.77\rho_w $, $z_1 \approx 18$\,\AA\, and $z_2 \approx 36$\,\AA.

The exponent $\sigma_{R}$ in Eq. in (13) presents the contribution
of capillary waves to the structure of the interface.
Its square $\sigma_{R}^2 \approx ( k_BT/2\pi\gamma )\ln(Q_{max}/Q_{min})$ is
specified by the short-wavelength limit in the spectrum
of thermal fluctuations of the interface $Q_{max} = 2\pi/a$ ($a\approx 10$\,{\AA}
is about the molecular radius)
and by the angular resolution of the detector
$Q_{min}=q_z\Delta\beta/2$ ($q_z = 0.05$ {\AA}$^{-1}$) \cite{Schwartz, Weeks, Braslau2,  SCH211, Tikh311, Tikh111}. The parameter $\sigma_{R}$ in the experiments varies from 4\,\AA\, to 6\,\AA.

\begin{figure}
\hspace{0.5in}
\epsfig{file=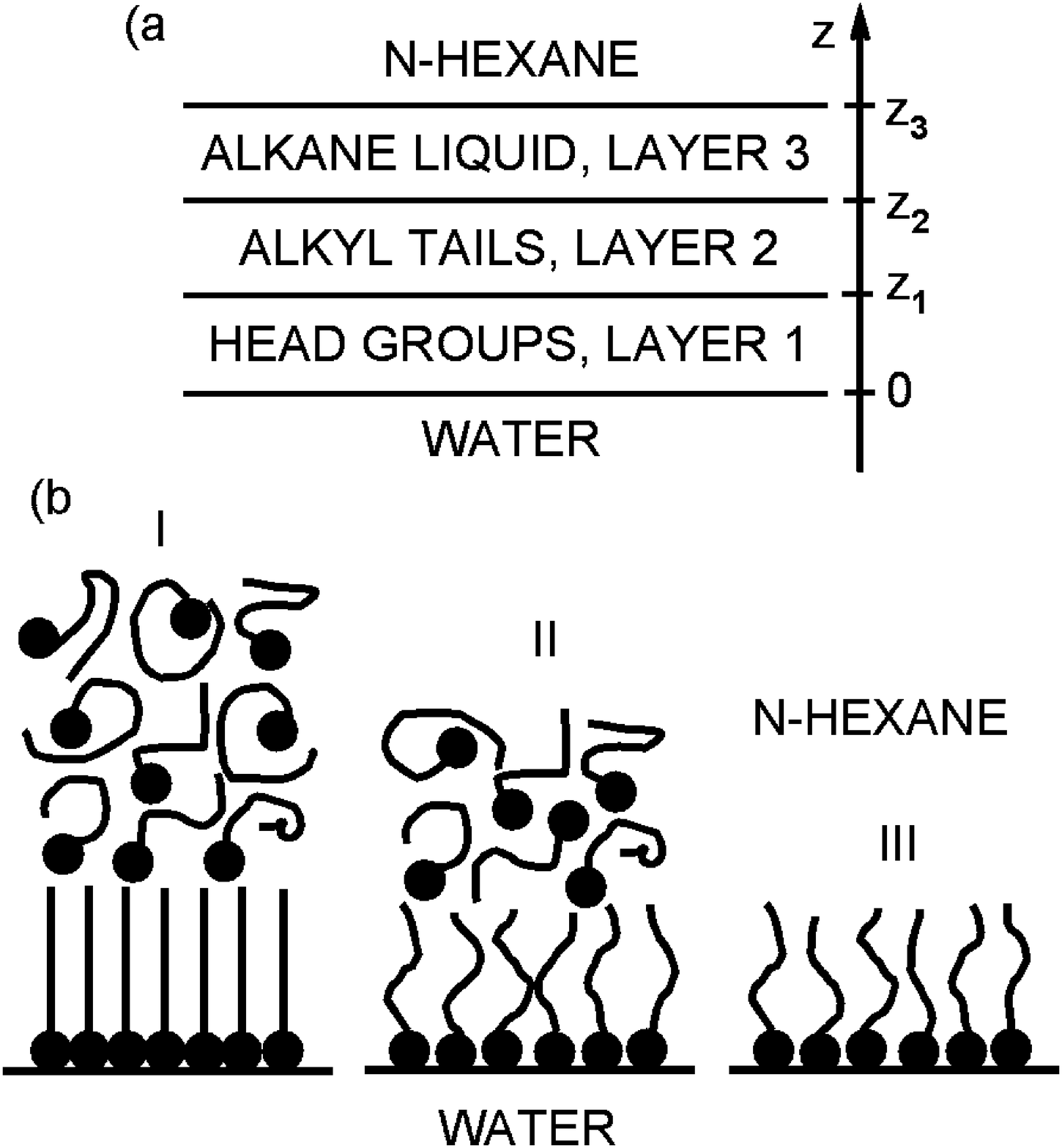, width=0.7\textwidth}

Figure 4. (a) Three-layer model of the adsorbed layer of
melissic acid C$_{30}$H$_{60}$O$_2$ at the n-hexane -- water interface.
(b) Transverse structure of the n-hexane -- water interface:
(I) the thick layer with a crystalline monolayer at $T<T_c$,
(II) the layer with a liquid monolayer in the intermediate
region $T_c<T<T^*$, and (III) the homogeneous liquid
monolayer at $T>T^*$.

\end{figure}

The intensities $I_n$ calculated with $\Phi(q)_m$ are shown
by solid lines in Figs. 2 and 3. The observed intensity
in the solid phase at $T < T_c$ is more than an order of
magnitude larger than the calculated value. At the
same time, the experimental data for the liquid phase
of the monolayer at $T > T^*$ are well described by
Eq. (13) without the variation of the parameters of the
layer. Thus, scattering at $T < T^*$ occurs on a structure
more complex than the homogeneous monolayer and
a transition from the solid phase of the surface to the
liquid monolayer occurs at two stages with the temperatures $T_c$ and $T^*$.

The simplest three-layer model that qualitatively
explains scattering data shown in Figs. 2 and 3 (dashed
lines) and simultaneously reflectometry data reported
in [3, 5] is parameterized by the structure factor of the
form (see Fig. 4à)
\begin{equation}
\displaystyle
\Phi(q)^*_m +\frac{ \displaystyle \delta\rho e^{-\sigma^2q_z^2/2 }}{ \displaystyle \Delta\rho }  e^{-iq_zz_3}.
\end{equation}
Here, the second term describes the third adsorbed
homogeneous layer with the thickness $z_3-z_2$ and
density $\rho_h + \delta\rho$, $\sigma$ is the width of the interface
between this layer and bulk of n-hexane, and $\Phi(q)^*_m$ is
specified by Eq. (13) with the change $\rho_3 \to \rho_h + \delta\rho$.
At $T > T^*$, the excess surface density vanishes: $\delta\rho(z_3-z_2)=0$.

The joint analysis of the data for $I_n$ and $R(q_z)$ with
the use of Eq. (14) shows that all parameters of the
layer at $T<T_c$ hardly depend on the temperature $T$.
Both nonspecular scattering at small $\beta$  values
($\sigma^2q_z^2<<1$) and reflectometry data are satisfactorily
described at the following parameters of the third
layer: $z_3-z_2 \sim 200$\,\AA, $\delta\rho \approx 0.1 \rho_w \div 0.25 \rho_w$,
and $\sigma \approx 10 \div 20$\,\AA{} Since the contribution to $R(q_z)$ from the
second term in Eq. (14) decreases rapidly with an
increase in $q_z$ and becomes negligibly small at $q_z> 0.075$\,\AA$^{-1}$,
the correction of the parameters of the
solid monolayer in $\Phi(q)^*_m$  is insignificant (within the
error). Finally, the existing data and the used approach
cannot provide a reliable determination of the parameters
of the possible internal structure of the third
layer.

The electron density in the third layer $\rho_h + \delta\rho$ at $T<T_c$
corresponds to a high-molecular-weight alkane liquid
\cite{Small}. The fraction of melissic acid in this layer is estimated
as $f=\delta\rho/(\rho_m - \rho_h) \approx 0.8$, where $\rho_m \approx 0.9 \rho_w$ is
the density of the liquid monolayer of C$_{30}$-acid.
According to Fig. 3, $\delta\rho(z_3-z_2)\to 0$ in the intermediate
region $T_c<T<T^*$ at $T \to T^*$. Unfortunately,
existing data are insufficient to obtain detailed information
on this asymptotic behavior.

The formation of a multilayer structure at the
alkane -- water interface was previously detected in
adsorbed layers of some monohydric alcohols and,
more recently, in layers of mixtures of fluorocarbon
alcohols \cite{TAMSCH, Takiue2}. For example, the low-temperature
phase of dodecanol adsorbed at the n-hexane -- water
interface is a high-molecular-weight alkane liquid whose density
is equal to the density of the layer of melissic acid
in the transitional region $T_c<T<T^*$. Nevertheless,
the two-dimensional evaporation phase transition in
dodecanol is fundamentally different from the melting
transition in the C$_{30}$-acid because it is described by
only a single critical temperature.

\begin{figure}
\hspace{0.5in}
\epsfig{file=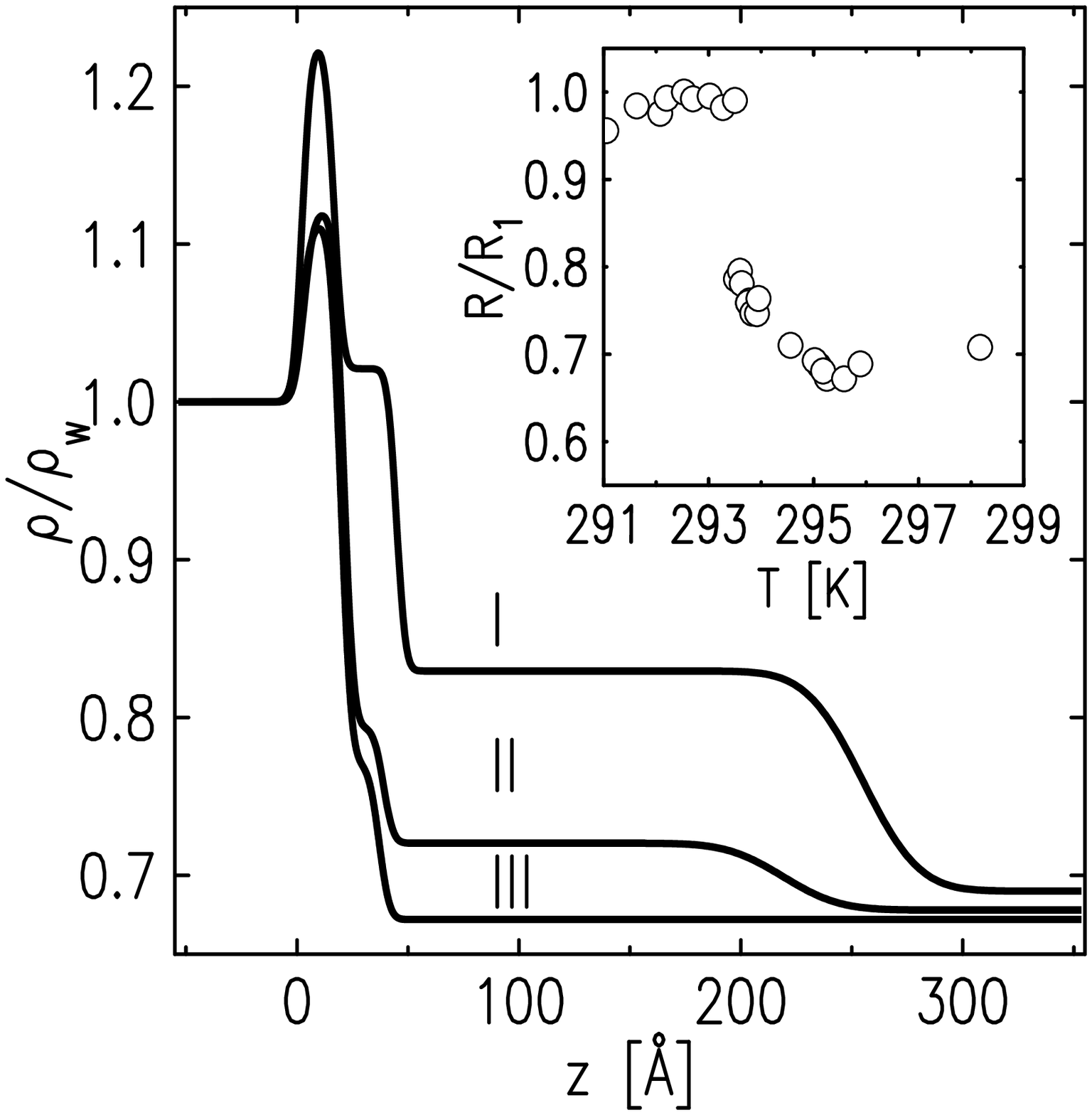, width=0.7\textwidth}

Figure 5. Model profiles of the electron density: (I) the three-layer
model with a solid monolayer $(T<T_c)$, (II) the three-layer
model in the intermediate region $(T_c<T<T^*)$, and
(III) the two-layer model of a liquid monolayer $(T>T^*)$.
The inset shows the temperature dependence of the normalized
reflection coefficient $R/R_1$ at $q_z=0.05$\,\AA$^{-1}$,
where $R_1$ is the reflection coefficient at $T \approx 292.2$\,K.

\end{figure}

Two-dimensional phase transitions in two stages
are characteristic of systems, e.g., with surface active
mixtures of fluorocarbon and hydrocarbon alcohols
\cite{Takiue1}. The existence of two critical temperatures was
mentioned in \cite{Tamam}, where the crystallization of monolayers
of cation surfactants CTAB and STAB was considered.
However, in both cases, the authors discussed
the structures of monolayers rather than extended
multilayer structures.

Finally, Fig. 5 shows profiles of the electron density
for surface structures (see Fig. 4b). The structure I at $T < T_c$
(Fig. 5) consists of a solid monolayer with a
thickness of $\approx 41$\,\AA{} and a layer of a high-molecular-weight
alkane liquid with a thickness of $\sim 200$\,\AA. With an
increase in the temperature, a sharp jump occurs in
the reflection coefficient at $T_c \approx 293.5$\,K (see the inset
in Fig. 5), which indicates the melting of the monolayer
of melissic acid immediately at the interface with
n-hexane (structure II, Fig. 5). With a further increase
in the temperature, $\delta\rho(z_3-z_2)\to 0$, which is accompanied
by a decrease in the diffuse scattering intensity.
At $T > T^*\approx 320$\,K, only the liquid monolayer of C$_{30}$-acid
with a thickness of $\approx 36$\,\AA{} remains at the interface
(structure III, Fig. 5). Such a behavior of the system
indicates that the two-dimensional crystallization
phase transition in the interface occurs with a decrease
in the temperature $T$ at $T_c$ after the wetting phase transition
at $T^*$.

I am grateful to Prof. Mark L. Shlossman and Prof. Vladimir I. Marchenko
for stimulating discussions of the experimental results.
This research used resources of the National Synchrotron Light Source, a U.S. Department of Energy
(DOE) Office of Science User Facility operated for the DOE Office of Science by Brookhaven National
Laboratory under Contract No. DE-AC02-98CH10886. Beamline X19C received support from the ChemMatCARS
National Synchrotron Resource, the University of Chicago, the University of Illinois at Chicago,
and Stony Brook University.

\end{document}